# Signatures of evanescent transport in ballistic suspended graphene-superconductor junctions

Piranavan Kumaravadivel[1*] and Xu Du[1**]

[1]Department of Physics and Astronomy, Stony Brook University, NY, 11794-3800, USA.

## Abstract

**In Dirac materials, the low energy excitations behave like ultra-relativistic massless particles with linear energy dispersion. A particularly intriguing phenomenon arises with the intrinsic charge transport behavior at the Dirac point where the charge density approaches zero. In graphene, a 2-D Dirac fermion system, it was predicted that charge transport near the Dirac point is carried by evanescent modes, resulting in unconventional "pseudo-diffusive" charge transport even in the absence of disorder. In the past decade, experimental observation of this phenomenon remained challenging due to the presence of strong disorder in graphene devices which limits the accessibility of the low carrier density regime close enough to the Dirac point. Here we report transport measurements on ballistic suspended graphene-Niobium Josephson weak links that demonstrate a transition from ballistic to pseudo-diffusive like evanescent transport below a carrier density of ~ $10^{10}$ cm$^{-2}$. Approaching the Dirac point, the sub-harmonic gap structures due to multiple Andreev reflections display a strong Fermi energy-dependence and become increasingly pronounced, while the normalized excess current through the superconductor-graphene interface decreases sharply. Our observations are in qualitative agreement with the longstanding theoretical prediction for the emergence of evanescent transport mediated pseudo-diffusive transport in graphene.**

*Present address: Yale University, Connecticut, USA.

**Corresponding email: xu.du@stonybrook.edu

**Introduction**

In the last decade, there has been an increasing interest in the transport properties of Dirac materials such as graphene, topological insulators, topological Dirac semimetals [1-6] and in 2D photonic crystals with a Dirac-like spectrum [7-9]. With a linear energy dispersion, these materials show drastically different electronic properties compared to the conventional massive fermion systems. One of the most interesting questions related to the Dirac-like spectrum is what happens to the charge transport at the charge neutral Dirac point, where zero Fermi energy co-exists with a zero energy gap/barrier. In graphene, a 2-dimensional (2D) massless Dirac fermionic system, this problem has been theoretically investigated using the standard Landauer formalism, where conductance is supported by the transmission of various transverse modes. Different from the conventional massive electron gas systems, the transmission of the transverse modes in graphene is governed by the Dirac-Weyl equation. This difference becomes crucial at the electron-hole degenerate Dirac point. Here the plane wave-like propagating modes that contribute to the conventional ballistic conductance at higher Fermi energies give way to evanescent modes. Evanescent modes are comprised of imaginary wave vectors and their transmission amplitudes decay with length. Therefore the conductance from evanescent modes becomes inversely proportional to the length of the graphene channel[10]. This characteristic is similar to conductance in *diffusive* systems. In graphene at the charge-neutral Dirac point, the evanescent mode conductivity reaches a quantum limited value of $\frac{4e^2}{\pi h}$ [11].

Evanescent transport at the charge neutral Dirac point is a phenomenon unique to Dirac materials. For graphene, several longstanding theoretical proposals have been made as to how pseudo-diffusive charge dynamics should manifest itself in transport characteristics, based on the nature and distribution of charge transmission in the conduction channels. For example, in a short and wide ballistic graphene strip, the fluctuations in electrical current (shot noise) are found to be enhanced and have strong energy dependence in the evanescent transport regime, a phenomenon analogous to the "*Zitterbewegung*" of relativistic particles[11].Transport in ballistic graphene (G)- superconductor (S) hybrid devices is also expected to show pseudo-diffusive signatures [12-14]. Of particular relevance to the work presented here, it was predicted that in short ballistic S-G-S junctions the quasiparticle current-voltage characteristics due to the superconducting proximity effect should display a strong energy-dependence near Dirac point, reflecting the crossover from ballistic to pseudo-diffusive charge carrier transmission. In realizing these predictions, the experimental work carried out so far[15-19], have been impeded by numerous technical challenges. To reach the energy scale required for evanescent transmission, the Fermi wavelength should be of the order of $\lambda_F = \frac{hv_F}{E_F} \sim 2\pi L$. Here $v_F \approx 10^6$ m/s is the energy independent Fermi velocity, $E_F$ is the Fermi energy and L is the length of the graphene channel. Therefore the potential fluctuations ($\delta E_F$) near the neutrality point (NP) should be small, a few meVs, even for a micrometer long channel. In addition, charge carrier scattering should largely be eliminated, so that the transmission of the carriers reflects the intrinsic nature of the transverse modes. This requires the devices to be ballistic. Due to the strong substrate-associated disorder, previous observations were marred by the presence of large potential fluctuations (for graphene on $SiO_2$, $\delta E_F$ typically ranges from 25 to 100meV) and short mean free path (usually <<100 nm).More recently, monolayer graphene/h-BN hetero-structures have demonstrated ballistic transport[20]. Josephson current has also been observed in these structures when coupled with superconductors[21-23].However, methods for achieving both very low carrier density and highly transparent S-G interfaces are still under progress. To the best of our knowledge, an unambiguous experimental study of evanescent transport in graphene has not been reported.

In this letter, we present our study on charge transport in ballistic suspended graphene-Niobium (Nb) superconducting weak links. When approaching the NP ($n<10^{10}$ cm$^{-2}$),where the evanescent mode transport starts to dominate over the conventional propagating modes, the multiple Andreev reflection related features known as sub-harmonic gap structures (SHGS) become more pronounced. We also find that the normalized excess current ($I_{exc}R_N$), which remains constant at high carrier densities, becomes suppressed rapidly at low carrier densities near the NP. Both observations are in contrast with previous experimental observations in disordered graphene superconductor junctions where both SHGS and $I_{exc}R_N$ show no significant gate dependence [19,24,25]. Our results are in qualitative agreement with the theoretical predictions and provide strong evidence for pseudo-diffusive transport in ballistic graphene.

**RESULTS**

**Device and device characteristics**

The devices used in this study are suspended graphene–Nb Josephson weak links fabricated on Si/SiO$_2$ substrates. The device structure is illustrated in Figure 1 and its fabrication technique is described in the *Method* section. For the device discussed in this report, the graphene channel was designed to have a large aspect ratio (*W/L*) ~9 with width *W* =5.5µm and length=0.6µm. Such geometry minimizes any effect from the edges of graphene and complies with the theoretical prescription $W/L \geq 4$ [10].

After cooling down to ~10K, the device was current-annealed through which Joule heating removes the surface contaminant from the fabrication process. The gating curves of the device after current annealing, measured at two different temperatures, 9K (~$T_c$) and 1.5K, are presented in figure 1b. At *T*~9K the device resistance shows a very strong and sharp gate-dependence. The excellent quality of the device is evident from the quantum Hall (QH) measurements. As shown in figure 1c, at *T*=1.5K, and in a low magnetic field of *B*= 300mT, pronounced magneto-oscillations are already observed. At *B*=500mT, the sample displays fully developed anomalous quantum hall plateaus at $\nu=\pm 2, \pm 6,...$ where $\nu = \frac{nh}{eB}$. From these QH plateaus, we find the carrier density (*n*)-gate voltage ($V_g$) relation: $n = 1.84 \times 10^{10} \times (V_g - V_{NP})$[Volt]cm$^{-2}$ where $V_{NP} = -0.9$V is the gate voltage at NP. This is consistent with the estimation using the geometrical capacitance considering 285nm SiO$_2$ in series with 220nm (thickness of the PMMA spacer) of vacuum. On the higher mobility electron side, we also observe additional oscillatory features in resistance, *R* ($V_g$), at $\nu$=1,4,8..., etc., which may be attributed to the onset of broken symmetry states. The resistance at the NP displays diverging behavior with increasing field, starting at a low *B*~0.3T. All the features observed here suggest that the sample is of extremely high quality, with long mean free path and minimal potential fluctuations.

Based on the carrier density dependence of the resistivity, we find for the device a maximum Hall mobility of >250,000 cm$^2$/Vs and a mean free path which is limited by the sample length. From the smear of the resistance near the NP and based on the gate voltage-carrier density relation obtained from the quantum Hall measurement, we estimate the minimum carrier density $n_s \sim 1.4 \times 10^9$ cm$^{-2}$. This corresponds to a potential fluctuation $\delta E_F = \hbar v_F \sqrt{n_s \pi} \sim 4.4$ meV at NP, the smallest value observed so far in superconductor-graphene devices. To characterize the Fermi wavelength, similar to reference 12, we use a dimensionless parameter $\kappa = \frac{E_F L}{\hbar v_F} = \frac{2\pi L}{\lambda_F}$. Therefore the maximum $\delta E_F \sim 4.4$ meV in our sample corresponds to $\kappa \sim 4.0$. The resistivity at NP is $\sim 19 \text{k}\Omega = 0.93 \frac{\pi h}{4e^2}$. The small discrepancy from the theoretical value of $\frac{\pi h}{4e^2}$ may be attributed to the presence of electron hole puddles and finite Coulomb scattering which, in practice, cannot be completely avoided.

**Differential resistance and IV characteristics at large carrier densities.**
Next, we study the superconducting proximity effect at $T=1.5$K in absence of magnetic field. Compared to $T>T_C$, the junction resistance is significantly reduced (Figure 1c). At large gate voltages far from the Dirac point, the resistance approaches zero with the development of a sizable supercurrent, as shown by the *I-V* curve in Figure 1c inset. The observed *I-V* characteristic is well-captured by the resistively and capacitively shunted Josephson junction (RCSJ) model in presence of thermal fluctuations[26], as indicated by the solid curve which fits quantitatively to the data. From the fitting and by using a back-gate-coupled source-drain capacitance of $C = 0.9$pF and an Andreev reflection reduced resistance of $R=100\Omega$, we estimate a critical current of $I_c = 220$nA. The strong impact of thermal fluctuations accounts for the absence of a real zero resistance state, especially near the Dirac point where critical current is low. Hence the Josephson energy is small compared to the thermal energy: $\frac{I_c \hbar}{2e} \leq k_B T$. Figure 1d shows the differential resistance $\frac{dV}{dI}$ as a function of bias voltage ($V_{bias}$) taken at $V_g - V_{NP} = 7.5$V ($\kappa \sim 39$). From the curve we obtain the superconducting gap at the SN interface ($\Delta$) ~0.34meV. This value is significantly smaller than the BCS gap of Nb ($\Delta_{BCS}(0) \sim 1.764 k_B T_c \sim 1.37 \text{meV}$) and varies slightly from sample to sample. Similar reduction has also been observed in superconductor-nanowire weak links[27,28] and can be attributed to the strong anti-proximity effect of the Ti/Pd buffer layer on the surface superconducting gap of Nb at the interface. Since the metal-metal resistance (both between surface and bulk Nb, and between surface Nb and the Ti/Pd buffer layer) is expected to be negligible, most of the bias voltage would drop across the graphene channel between the two graphene-metal interfaces, where graphene "sees" an anti-proximity-reduced gap of the superconductor. Such gap reduction may be minimized by experimenting with different buffer layer materials, as demonstrated by the recent works[22]. For the device presented here, at finite $V_{bias} \ll 2\Delta$, the resistance drops down to ~40% of the normal resistance ($R_N$) due to the Andreev reflection process. Evaluating the value of the normalized excess current $I_{exc}R_N \sim 0.4\text{mV} \sim \Delta/e$ and using the OTBK model[29] we estimate the dimensionless barrier strength of the interface, $Z \sim 0.5$. The SHGS are very weak, consistent with the theoretical prediction for ballistic channels with high transmission[30,31]. Other factors that may affect the weak SHGS include the reduced interfacial superconducting gap and the relatively high measurement temperature.

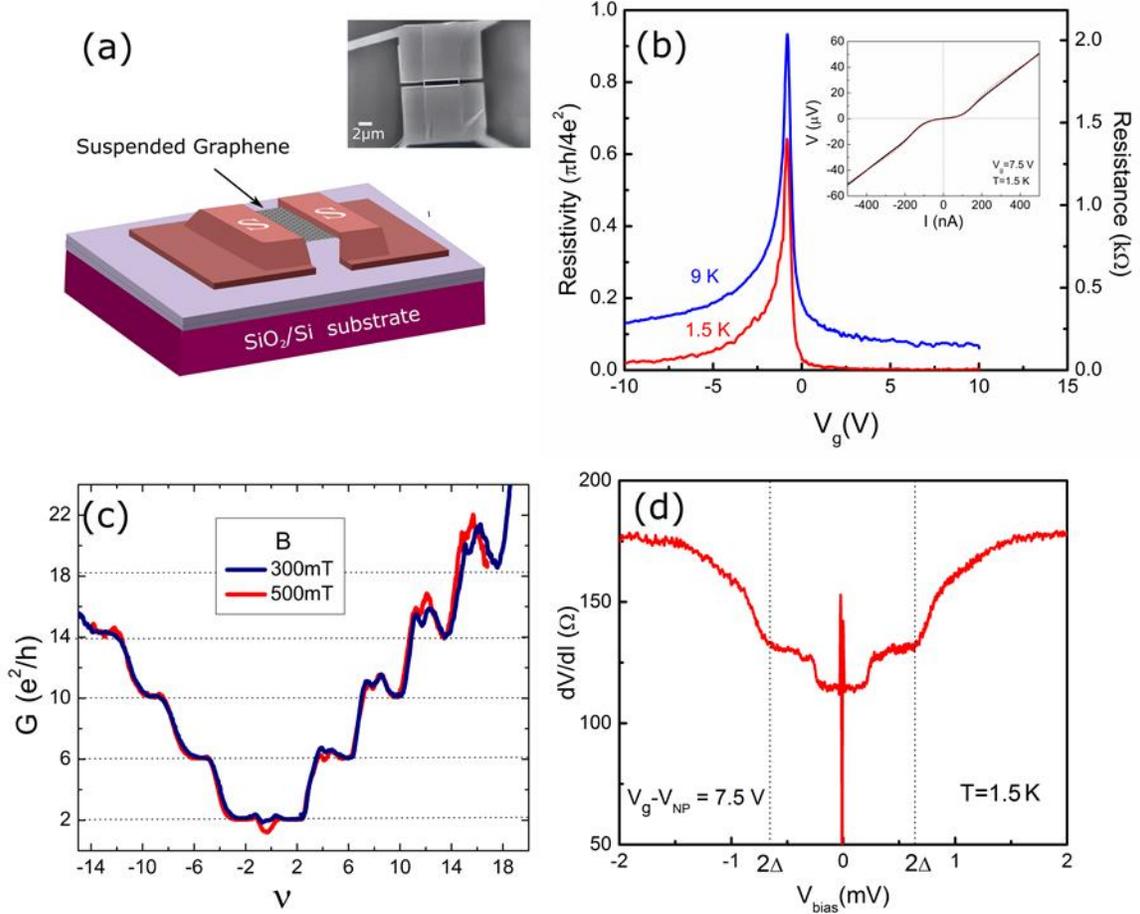

**Figure 1.** (Color online) **Device characteristics:** (a) Main panel: Device schematics. Inset: SEM Image of the device. Scale bar is 2 μm. The graphene channel is highlighted by the open rectangle (b) Resistivity in units of ($\pi h/4e^2$) as a function of gate voltage ($V_g$) at $T$= 9 K~$T_c$ (blue) and T =1.5 K (red). Inset: *I-V* characteristic showing development of supercurrent. The red dotted curve is experimental data measured for $Vg$ = 7.5 V; the black solid curve is a fitting to the data using the RCSJ model with thermal fluctuations. (c) Quantum Hall measurements: Conductance versus filling factor for two different magnetic fields 300 mT (blue) and 500 mT (red). (d). Differential resistance as a function of bias voltage ($V_{bias}$) at T=1.5K. Gate voltage is 7.5V away from the NP gate voltage ($V_{NP}$). The corresponding κ ~39(see text) and the induced gap ($\Delta$) =0.34 meV.

**Differential resistance near the Dirac point**

Now we focus on the behavior of the differential resistance as we approach the NP. Figure 2a shows the normalized differential resistance $\frac{1}{R_N}\frac{dV}{dI}$ as a function of $V_{bias}$ obtained at various gate voltages. While the (x-axis) bias voltage values of the SHGS remain gate-independent, significant gate-modulation of the (y-axis) line-shape can be observed. When $V_g - V_{NP} < 0.6\text{V}$ ( $n < 10^{10}\text{cm}^{-2}$ and $\kappa < 9$ ), the differential resistance curve starts to develop a pronounced dip at $V_{bias} = 2\Delta/e \sim 0.68\text{mV}$. As the $V_g$ is ramped further towards the NP, i.e., for $V_g - V_{NP} < 0.3\text{V}$ ( $n < 5.7 \times 10^9 \text{cm}^{-2}$ and $\kappa < 5.5$ ), while the dip at $V_{bias} = 2\Delta/e$ continues to be deeper, other SHGS start to emerge at low $V_{bias}$. All the observed features appear at $V_{bias} = \pm 2\Delta/ne$ where *n=1, 2, 3…* as expected for multiple Andreev reflection processes and their positions in $V_{bias}$ are independent of $V_g$. The higher order features are within noise. The observed features are sharpest at NP where the SHGS at *n=1, 2 and 3* are all easily resolvable. With $V_g$ ramped across the NP and deeper into the hole side (κ= -1.9), the SHGS begin to weaken again.

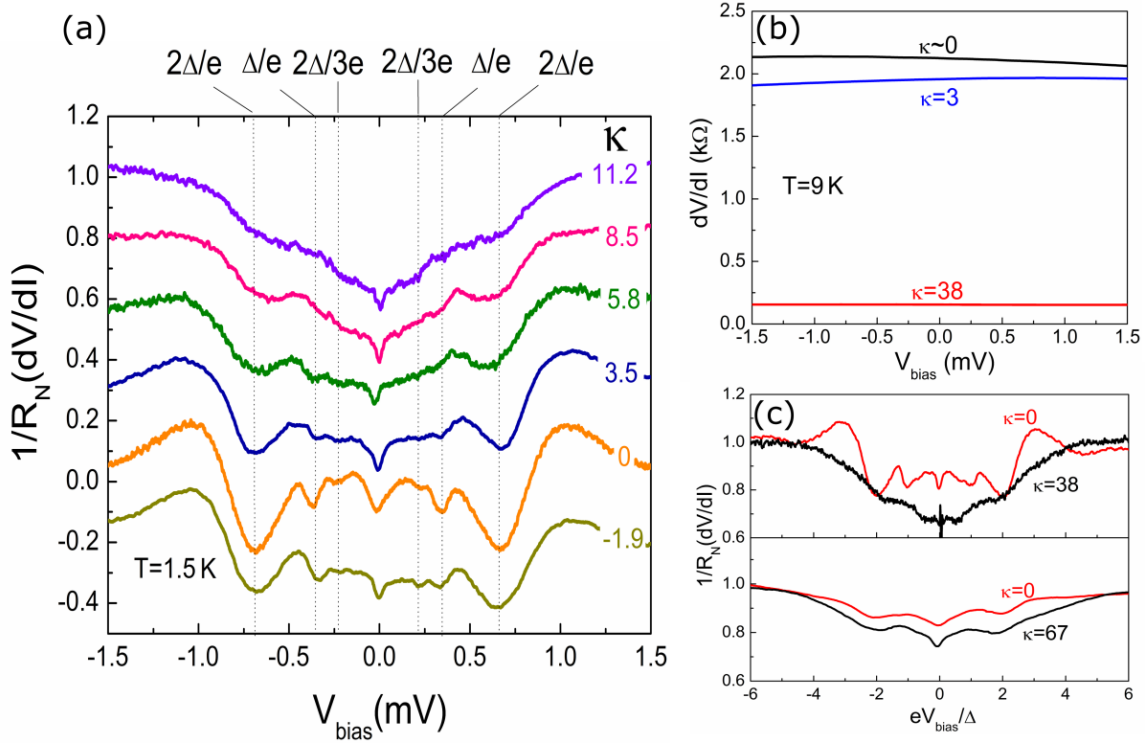

**Figure 2**. (Color online) **Differential resistance**. **(a)** Normalized differential resistance $\frac{1}{R_N}\frac{dV}{dI}$ versus bias Voltage ($V_{bias}$) for different $\kappa = \frac{E_F L}{\hbar v_F}$ for the ballistic S-G-S- device: Ti/Pd/Nb contacts at T=1.5K. Individual curves are shifted for clarity. Dotted lines indicate SHGS at $V_{bias}=\pm 2\Delta/ne$ for $n$=1, 2 and 3 where 2Δ=0.68meV. **(b)** The differential resistance (not normalized) of the ballistic S-G-S device measured slightly below the superconducting transition temperature: $T$ = 9K. **(c)** A direct comparison between the ballistic S-G-S device (upper panel) with a typical diffusive device with the same Ti/Pd/Nb contacts (lower panel), for $\frac{1}{R_N}\frac{dV}{dI}$ curves taken at the charge neutrality point and at large gate voltages. Both measurements are at T=1.5K.

The dip at $n=1$ shows the most prominent response to the gate voltage. In conjunction with the appearance of the pronounced SHGS, the overall shape of the normalized differential resistance curve transforms from a "V"-shape to a shallower profile. We note that the observed gate-dependence of $\frac{dV}{dI}$ versus $V_{bias}$ curves is specific to superconductivity. As shown in Figure 2b, above $T_c$ the "background" $\frac{dV}{dI}$ versus $V_{bias}$, within the bias voltage range studied here, shows roughly no curvature.

For a comparison with the ballistic device (Fig.2c,upper panel), we present data of normalized differential resistance vs $eV_{bias}/\Delta$ close to and away from NP of a diffusive device with similar Ti/Pd/Nb contacts (Fig.2c,lower panel). Here the graphene channel sits on SiO$_2$ and has a mean free path $l_{mfp} \sim 30\text{nm} \ll L$ and $\delta E_F \sim 40\text{meV}$. Other than the dip near zero-bias that is associated with the precursor of supercurrent, the gate voltage dependence of the SHGS for the disordered devices is much less significant compared to the ballistic devices. The quantitative variation in SGHS between the neutrality point and at large doping is not unexpected: even with strong potential fluctuations, charge neutrality can still be reached in patches of the channel when the device resistance is gate-tuned to be close to the maximum. Hence one should still expect a finite contribution of the evanescent transport near the Dirac point[24]. On the other hand, because the neutral patches are always surrounded by electron and hole "puddles", their impact on device transport gets severely smeared, and the gate-dependence vastly broadened. For devices with extremely large potential fluctuations (i.e., those with very broad $R$ vs. $Vg$ dependence around the Dirac point), indeed, the SGHS becomes almost completely gate-independent as demonstrated in previous work with diffusive graphene Josephson links with Al/Ti contacts[19].

To understand the observed gate-dependent SHGS, we consider Fermi energy modulation of charge transmission in a ballistic graphene device. In general, the transmission of Dirac electrons in graphene can be described by a summation of contributions from the boundary-defined transverse modes, each satisfying the Dirac-Weyl equation. Depending on the Fermi energy, the lowest $N(E_F) = \frac{E_F W}{\pi \hbar v_F}$ modes are propagating with real wave vectors, while the higher $N > N(E_F)$ modes are evanescent with imaginary wave vectors[14]. At high densities, the channels in the ballistic graphene strip are propagating with high transmission. Approaching the NP, however, the propagating modes become suppressed and the evanescent modes contribute increasingly to the conduction with more channels having a lower transmission. This change in transmission distribution is the underlying reason for the emergence of pseudo-diffusive transport signatures. With superconducting contacts, it has been shown that the oscillatory amplitude of the SHGS is due to the contribution from the low-transmission channels[12,30]. As a result, close to the NP where evanescent modes with low transmitting probability dominate, the SHGS are more pronounced than at large gate voltages. The stronger gate-dependence of lower order

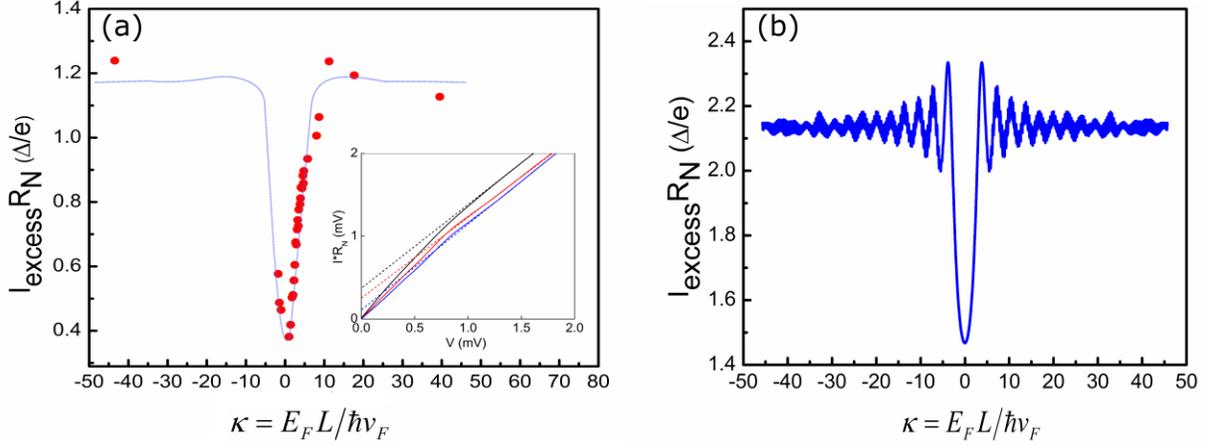

**Figure 3.** (Color online) **Normalized Excess current ($I_{exc}R_N$) as a function of** $\kappa = \dfrac{E_F L}{\hbar v_F}$ **a.** Experimental data: $I_{exc}R_N$ is calculated from the current-voltage characteristics of the device for different $V_g$ at 1.5K. The (blue) line is drawn as a guide to the eye. Inset: determination of excess current through extrapolation of $I$-$V$ curves. The three $I$-$V$ curves are taken at $\kappa = 0$ (blue), 4.6 (red), and 38 (black). Note that the characteristics associated with supercurrent $I_c R_N$, are too small to resolve over the range of data presented here, even for $\kappa = 38$. **b.** Theoretical calculation: for short ballistic S-G-S junction with W=5.5μm and L=0.6μm at zero temperature. The calculation is performed using the model described in ref. 12.

SHGS, especially $n=1$, is also consistent with the theory. This is because, compared to the higher order SHGS, the lower order ones are formed by Andreev quasi-particles that traverse the graphene channel fewer times and hence involves more contribution from the low-transmission channels.

**Normalized excess current**

Further comparison with theoretical predictions for pseudo-diffusive transport can be made by characterizing the excess current in the S-G-S device. Generally in a SNS junction when $eV_{bias} \gg 2\Delta$, the current though the sample consists of a normal 'Ohmic' current ($I_N$) and an 'excess' current ($I_{exc}$) due to the superconducting proximity effect. Compared to the Josephson current, excess current is much more robust against the influence of the electromagnetic environment, and hence provides a reliable parameter for characterizing the proximity effect. In S-G-S junctions the abundance of nearly ballistic modes at large Fermi energies leads to large excess current. However, in the vicinity of the NP the number of highly transmitting propagating modes decreases and the charge transport becomes increasingly evanescent thereby decreasing the excess current. Figure 3a shows the normalized excess current $I_{exc}R_N$ in our device (extracted from current-voltage curves at various gate voltages as shown in Figure 3a inset) as a function of κ. For κ>9, i.e., in the ballistic transport regime $I_{exc}R_N$~0.4mV~1.2Δ/e whereas for κ <9, there is a clear gate dependence. The excess current sharply reduces when approaching the NP. This reduction coincides with the onset of the enhanced SHGS as shown in figure 2a. For short ($\frac{\hbar v_F}{L} \gg \Delta$) ballistic graphene Josephson junctions, the excess current has been theoretically studied[12]. For $L$=0.6μm, $\frac{\hbar v_F}{L} \sim$ 1meV and since the induced gap $\Delta \sim$ 0.34meV our device marginally satisfies the short junction limit. The observed gate modulation of normalized excess current is in qualitative agreement with the zero temperature theoretical calculations shown in Figure 3b. Here we use our experimental device geometry ($W$=5.5μm and $L$=0.6μm) and following previous theoretical work[12], the total excess current in a short Josephson junction is calculated as a sum of its individual contribution from all the transverse modes[12]:

$$I_{exc} = \sum_n \frac{2e\Delta}{h} \frac{\tau_n^2}{1-\tau_n}\left(1 - \frac{\tau_n^2}{2(2-\tau_n)\sqrt{1-\tau_n}} \ln\left(\frac{1+\sqrt{1-\tau_n}}{1-\sqrt{1-\tau_n}}\right)\right)$$

, where $\tau_n$ is the transmission probability of the n$^{th}$ transverse mode at the Fermi level:

$$\tau_n = \left|\frac{2\delta^2 - 2(k_n - q_n)^2}{e^{k_n L}(q_n - k_n + i\delta)^2 + e^{-k_n L}(q_n - k_n - i\delta)^2}\right|.$$

Here the transverse momentum of the n$^{th}$ mode $q_n = \frac{\pi}{W}\left(n+\frac{1}{2}\right)$, $\delta = \frac{E_F}{\hbar v_F}$ and $k_n = \sqrt{q_n^2 - \delta^2}$ and $n$=0,1,2,... labels the modes. The discrepancy between the theory and our observation, especially in the values for normalized excess current, may be attributed to the following factors. First, the theory assumes an ideal SN interface (Z=0), whereas in our device Z~0.5. Secondly, our measurements were carried out at a base temperature of ~ 1.5K, while the theory does not consider a finite temperature. Both these factors contribute to the reduction of the normalized excess current. In addition, for lowest values of κ, the theory does not consider the presence of the electron hole puddles thatexists in an actual device. This and due to the smearing from finite temperature, we observe a slight broadening in the excess current dip with an onset at κ~9 compared to the theory (κ~4).

**DISCUSSION**

Besides evanescent transport, we also consider other possible mechanisms for the observed sharp reduction of excess current near the Dirac point. One possible scenario in S-G junctions which may happen at the Dirac point is specular Andreev reflection (SAR)[32,33]. SAR happens when the Fermi energy of graphene is less than the superconducting gap. However, the Fermi energy broadening in the present sample (~5meV) is much larger than the superconducting gap (<1meV). As a result SAR is completely smeared by the potential fluctuations. This is supported by the fact that we do not observe a sub-gap differential resistance peak whose bias voltage shifts with changing gate voltage/Fermi energy, which is the key signature of SAR.

We also consider the possible impact of an imperfect S-G interface that may also result in Fermi energy modification of transmission coefficients without any direct consequence of the Dirac fermionic nature of graphene. A careful study of our Nb/Pd/Ti-graphene contacts has been carried out using non-suspended devices with different channel aspect ratios. These measurements reveal a contacts resistance which is at least 10 times lower that the two-terminal resistance measured in this work, throughout the gate-tunable range. Based on Landauer formalism this indicates that negligible reflection happens at the superconductor-graphene interface due to the presence of "classical" contact resistance.

Another potential complication arises from the doping of the metal contacts which can extend into graphene, forming a *p-n* junction that imposes the charge carrier reflections [34,35]. A direct evidence of the presence of such interfacial *p-n* junction is the electron-hole asymmetry in the *R* vs. $V_g$ dependence. As the gate voltage is swept across the NP, the S-G interface changes from *p-n* to *n-n* and the asymmetry in the *R* vs. $V_g$ dependence can be associated with the transmission probability across the S-G junction. While the presence of such junctions may affect the SHGS and $I_{exc}$, we expect its gate voltage dependence to be gradual with no particular energy scale. In addition, such contact-doping associated reflection should give rise to asymmetry in the quasi-particle current-voltage characteristics that persists up to large gate voltages on both the electron and hole sides[22]. These are apparently not consistent with the observation of a sharp dip on the $I_{exc}R_N$ vs. $E_F$ dependence for $E_F<8$meV, and the qualitatively symmetric behavior with respect to the NP.

In summary, we have studied the transport properties of Dirac electrons extremely close to the charge neutrality point in ultrahigh quality superconductivity induced suspended graphene. We observe a clear transition from ballistic to pseudo-diffusive behavior of Dirac electrons. This transition is reflected in the Fermi energy dependence of sub-harmonic gap structures and excess current in the vicinity of the neutrality point. Our results unambiguously demonstrate long standing theoretical predictions for the emergence of evanescent transport mediated pseudo-diffusive transport in graphene. The devices developed in this work also opens up the opportunities for future studies of low charge density physics of Dirac electrons, such as electron-electron interactions and microwave/photo-responses very close to the Dirac point, where correlated and non-Fermi liquid behaviors may become strong.

## METHODS

**Device fabrication and electrical measurements**

For fabrication of the suspended graphene devices we use the method similar to work presented by Mizuno et.al[26]. The highly simple, robust and versatile technique requires no etchants. An illustration of the device is shown in Figure 1. The fabrication is carried out by exfoliating HOPG flakes on a polymethyl methacrylate (PMMA) spacer supported by a $SiO_2$/Si substrate. Here the PMMA is spin coated and hot baked at $180^0$C for 90s to form ~220nm thick layer. An imaging resist layer of the copolymer methyl methacrylate (MMA) is then spin-coated at 3000 rpm on top of the graphene/PMMA stack and baked at$150^0$C, for definition of the electrical contacts using electron beam lithography. By controlling the electron beam exposure dose, we develop the 3-dimensional surface profile necessary to suspend and support the graphene channel over the SiO$_2$/Si substrate. After e-beam lithography, the sample is exposed to UV ozone for 1.2 minutes before loaded into a metallization chamber with a base pressure below 1x 10$^{-8}$Torr. For electrical contacts, the sample is metalized with the buffer layers Ti and Pd, each of thickness ~1.0nm. Without breaking the vacuum, ~60 nm of the superconducting Nb film ($T_c$~9K, $Hc_2$~3.5 T) is then DC Magnetron sputtered on top of the buffer layers in Argon plasma, at a rate of ~ 1nm/s. The sputtering parameters and thickness of the metal films are determined to reduce the stress of the sputtered film on graphene and to establish a transparent interface. The lift off process is performed in two successive baths of hot (80$^0$C) acetone and isopropyl alcohol. The sample remains in the liquids during the process, and is finally transferred to the low surface tension hot (~60$^0$C) hexane[36] and then directly taken out and dried in air.

All the measurements were performed in an Oxford Instruments Variable Temperature Insert (VTI). To minimize microwave heating and lower the electron temperature, electromagnetic radiation is filtered at various stages, including a filter assembly of ferrite beads, RF chokes and pi-filters at room temperature, and cryogenic two-stage RC filters (1000Hz cut-off frequency) at 4.2K. For differential resistance measurements, a Keithley 6221 current source is used to supply a 10nA AC current while ramping the DC offset current. The AC voltage response is measured with a Stanford Research SR830 lock-in amplifier, while the DC bias voltage is measured using a Keithley 2182A nanovoltmeter.

**Numerical simulations on the RCSJ model**

We describe the Josephson switching behaviors of the graphene-Nb devices using the resistively and capacitively shunted Josephson junction (RCSJ) model. For a current-driven device, the total current through the junction splits into the "pure" Josephson junction, a "normal state" resistor and a shunt capacitor: $i = \sin\phi + \frac{\hbar}{2eRI_c}\frac{d\phi}{dt} + \frac{\hbar C}{2eI_c}\frac{d^2\phi}{dt^2}$.

Here, $i = I/I_c = (I_{app} + I_{noise})/I_c$ consists of an applied current and a noise current. $\phi$ is the macroscopic phase difference between the two superconducting leads; $I_c$ is the critical current in the "pure" Josephson junction; $R$ is the normal resistance at which is takes into account the Andreev reflection effect; and $C$ is the effective capacitance between the superconducting leads, coupled by the conducting backgate.

To solve the equation numerically, we change the differential equation to the difference equation: $\phi_{n+1} = \frac{1}{B}(i - \sin\phi_n)\Delta t^2 - \frac{A}{B}(\phi_n - \phi_{n-1})\Delta t + 2\phi_n - \phi_{n-1}$

Here $\Delta t$ is chosen to be much (~1000 times) smaller than the period of the Josephson oscillations. From $\phi(t)$ we can calculate the averaged DC voltage: $\langle V \rangle = \frac{\hbar}{2e}\left\langle \frac{d\phi}{dt} \right\rangle$. For $\phi(t)$ we used the Johnson noise current generated as a Gaussian white noise and related to the temperature by $\langle I_{\text{noise}}^2 \rangle = \frac{4k_B T}{R} f$, where $f$ is the bandwidth used in our simulation. For each generated Johnson noise, we calculate an IV curve based on the above method. The IV curves are averaged 100 times over randomized Johnson noise currents. The final results are compared with the experimental data.

More details of the RCSJ model simulation including its validation and the impact of damping, thermal fluctuation and skewed current-phase relation are reported in our previous work on the study of ballistic supercurrent in graphene[26].

**References:**

1   Geim, A. K. & Novoselov, K. S. The rise of graphene. *Nat. Mater.* **6**, 183-191(2007).
2   Castro Neto, A. H., Guinea, F., Peres, N. M. R., Novoselov, K. S. & Geim, A. K. The electronic properties of graphene. *Rev. Mod. Phys.* **81**, 109-162(2009).
3   Hasan, M. Z. & Kane, C. L. Colloquium : Topological insulators. *Rev. Mod. Phys.* **82**, 3045-3067 (2010).
4   Moore, J. E. The birth of topological insulators. *Nature* **464**, 194-198(2010).
5   Young, S. M. *et al.* Dirac semimetal in three dimensions. *Phys. Rev. Lett.* **108**, 140405(2012).
6   Liu, Z. K. *et al.* Discovery of a three-dimensional topological Dirac semimetal, Na3Bi. *Science* **343**, 864-867(2014).
7   Sepkhanov, R. A., Bazaliy, Y. B. & Beenakker, C. W. J. Extremal transmission at the Dirac point of a photonic band structure. *Phys. Rev. A* **75**, 063813(2007).
8   Zhang, X. D. Observing zitterbewegung for photons near the dirac point of a two-dimensional photonic crystal. *Phys. Rev. Lett.* **100**, 113903 (2008).

9   Zandbergen, S. R. & de Dood, M. J. Experimental observation of strong edge effects on the pseudodiffusive transport of light in photonic graphene. *Phys. Rev. Lett.* **104**, 043903 (2010).
10  Tworzydlo, J., Trauzettel, B., Titov, M., Rycerz, A. & Beenakker, C. W. J. Sub-Poissonian shot noise in graphene. *Phys. Rev. Lett.* **96**, 246802 (2006).
11  Katsnelson, M. I. Zitterbewegung, chirality, and minimal conductivity in graphene. *Eur. Phys. J. B* **51**, 157-160 (2006).
12  Cuevas, J. C. & Yeyati, A. L. Subharmonic gap structure in short ballistic graphene junctions. *Phys. Rev. B.* **74**, 180501(R)(2006).
13  Akhmerov, A. R. & Beenakker, C. W. J. Pseudodiffusive conduction at the Dirac point of a normal-superconductor junction in graphene. *Phys. Rev. B.* **75**, 045426 (2007).
14  Titov, M. & Beenakker, C. W. J. Josephson effect in ballistic graphene. *Phys. Rev. B.* **74**, 041401(R) (2006).
15  DiCarlo, L., Williams, J. R., Zhang, Y. M., McClure, D. T. & Marcus, C. M. Shot noise in graphene. *Phys. Rev. Lett.* **100**, 156801 (2008).
16  Danneau, R. *et al.* Shot noise in ballistic graphene. *Phys. Rev. Lett.* **100**, 196802 (2008).
17  Heersche, H. B., Jarillo-Herrero, P., Oostinga, J. B., Vandersypen, L. M. & Morpurgo, A. F. Bipolar supercurrent in graphene. *Nature* **446**, 56-59(2007).
18  Miao, F. *et al.* Phase-coherent transport in graphene quantum billiards. *Science* **317**, 1530-1533 (2007).
19  Du, X., Skachko, I. & Andrei, E. Y. Josephson current and multiple Andreev reflections in graphene SNS junctions. *Phys. Rev. B* **77**, 184507 (2008).
20  Wang, L. *et al.* One-dimensional electrical contact to a two-dimensional material. *Science* **342**, 614-617(2013).
21  Calado, V. E. *et al.* Ballistic Josephson junctions in edge-contacted graphene. *Nat Nanotechnol.* **10**, 761-764 (2015).
22  Ben Shalom, M. *et al.* Quantum oscillations of the critical current and high-field superconducting proximity in ballistic graphene. *Nat. Phys.* **advance online publication**, doi:10.1038/nphys3592(2015).
23  Allen, M. T. *et al.* Spatially resolved edge currents and guided-wave electronic states in graphene. *Nat. Phys.* **12**, 128-133(2016).
24  Heersche, H. B., Jarillo-Herrero, P., Oostinga, J. B., Vandersypen, L. M. K. & Morpurgo, A. F. Induced superconductivity in graphene. *Solid State Commun.* **143**, 72-76 (2007).
25  Deon, F., Sopic, S. & Morpurgo, A. F. Tuning the influence of microscopic decoherence on the superconducting proximity effect in a graphene Andreev interferometer. *Phys. Rev. Lett.* **112**, 126803(2014).
26  Mizuno, N., Nielsen, B. & Du, X. Ballistic-like supercurrent in suspended graphene Josephson weak links. *Nat. Commun.* **4**, 2716 (2013).
27  Mourik, V. *et al.* Signatures of Majorana fermions in hybrid superconductor-semiconductor nanowire devices. *Science* **336**, 1003-1007(2012).
28  Das, A. *et al.* Zero-bias peaks and splitting in an Al-InAs nanowire topological superconductor as a signature of Majorana fermions. *Nat. Phys.* **8**, 887-895 (2012).

29  Flensberg, K., Hansen, J. B. & Octavio, M. Subharmonic energy-gap structure in superconducting weak links. *Phys. Rev. B.* **38**, 8707-8711 (1988).
30  Bardas, A. & Averin, D. V. Electron transport in mesoscopic disordered superconductor-normal-metal-superconductor junctions. *Phys. Rev. B.* **56**, R8518-R8521 (1997).
31  Averin, D. & Bardas, A. ac Josephson Effect in a Single Quantum Channel. *Phys. Rev. Lett.* **75**, 1831-1834(1995).
32  Efetov, D. K. *et al.* Specular interband Andreev reflections at van der Waals interfaces between graphene and NbSe2. *Nat. Phys.* **advance online publication**, doi:10.1038/nphys3583(2015).
33  Beenakker, C. W. Specular Andreev reflection in graphene. *Phys. Rev. Lett.* **97**, 067007 (2006).
34  Blake, P. *et al.* Influence of metal contacts and charge inhomogeneity on transport properties of graphene near the neutrality point. *Solid State Commun.* **149**, 1068-1071 (2009).
35  Hannes, W. R., Jonson, M. & Titov, M. Electron-hole asymmetry in two-terminal graphene devices. *Phys. Rev. B.* **84**, 045414 (2011).
36  Tombros, N. *et al.* Large yield production of high mobility freely suspended graphene electronic devices on a polydimethylglutarimide based organic polymer. *J. Appl. Phys.* **109,** 093702 (2011).

**Acknowledgements**

The authors thank Dmitri Averin for insightful discussions, Laszlo Mihaly for support with cryogenic facilities, and Peter Stephens for providing single crystal HOPG. This work was supported by AFOSR under grant FA9550-14-1-0405.

**Contributions**

P.K. contributed to all aspects of the work including device fabrication, measurements and data analysis. X.D contributed to data analysis and supervised the experiment. The authors equally contributed to discussion of results and the writing of the manuscript.

**Competing interests**

The authors declare no competing financial interests.